# Intertwined polar, chiral, and ferro-rotational orders in a rotation-only insulator

Weizhe Zhang[1], June Ho Yeo[1], Xiaoyu Guo[1], Tony Chiang[2], Nishkarsh Agarwal[2], John T. Heron[2], Kai Sun[1], Junjie Yang[3], Sang-Wook Cheong[4], Youngjun Ahn[1,5,+], Liuyan Zhao[1,+]

[1]Department of Physics, University of Michigan, Ann Arbor, MI, USA
[2]Department of Materials Science and Engineering, University of Michigan, Ann Arbor, MI, USA
[3]Department of Physics, New Jersey Institute of Technology, Newark, NJ, USA
[4]Keck Center for Quantum Magnetism and Department of Physics and Astronomy, Rutgers Center for Emergent Materials, Rutgers University, Piscataway, NJ, USA
[5]Materials Science Division, Argonne National Laboratory, Lemont, IL, USA

[+]Corresponding author: Liuyan Zhao (lyzhao@umich.edu) & Youngjun Ahn (youngjun.ahn@anl.gov)

**Abstract**

Intertwined orders refer to strongly coupled and mutually dependent orders that coexist in correlated electron systems, often underpinning key physical properties of the host materials. Among them, polar, chiral, and ferro-rotational orders have been theoretically known to form a closed set of intertwined orders. However, experimental investigation into their mutual coupling and physical consequences has remained elusive. In this work, we employ the polar-chiral insulator $Ni_3TeO_6$ as a platform and utilize a multimodal optical approach to directly probe and reveal the intertwining among polarity, chirality, and ferro-rotational order. We demonstrate how their coupling governs the formation of domains and dictates the nature of domain walls. Within the domains, we identify spatial inversion symmetry as the operation connecting two domain states of opposite polarity and chirality, with a common ferro-rotational state serving as the prerequisite for these interlocked configurations. At the domain walls, we observe a pronounced enhancement of in-plane polarization accompanied by a suppression of chirality. By combining with Ginzburg-Landau theory within the framework of a pre-existing ferro-rotational background, we uncover the emergence of mixed Néel- and Bloch-type domain walls. Our findings highlight the critical role of intertwined orders in defining domain and domain wall characteristics and open pathways for domain switching and domain wall control via intertwined order parameters.

Polarity refers to directionality, while chirality denotes handedness[1–3]. Polar and/or chiral structures are ubiquitous in nature[4–8], spanning from fundamental particles to complex minerals and across physical, chemical, and biological systems. Remarkably, when polarity and chirality coexist within a single system, a third type of order, ferro-rotational order[9,10], naturally arises as a composite of the two. This ferro-rotational order, together with polar and chiral orders, form a closed set of intertwined orders: the presence of any two of these three orders necessitates the existence of the third. Unlike polarity and chirality, ferro-rotational order is a relatively recent theoretical concept[10,11] and has only recently been realized experimentally[12–17]. It is pictorially represented by a closed-loop, head-to-tail arrangement of electric dipole moments, resulting in zero net polarization and preserved inversion symmetry, but with a defined handedness. The order parameter characterizing this state is an axial vector moment $\vec{A}$, defined as the sum over cross products between the local dipole moments $\vec{P}_i$ and their respective position vectors $\vec{r}_i$, i.e., $\vec{A} = \sum_{i=1}^{N} \vec{r}_i \times \vec{P}_i$. Despite the conceptual link among polar, chiral, and ferro-rotational orders, direct experimental investigation into their interplay remains largely unexplored to date.

### Crystal structure, symmetry-breaking pathways, and a multimodal optical platform for Ni$_3$TeO$_6$

In this study, we use Ni$_3$TeO$_6$ as the material platform to investigate the intertwined relationship among polar, chiral, and ferro-rotational orders within structural domains and at domain walls. The crystal structure of Ni$_3$TeO$_6$ belongs to the polar-chiral point group 3, characterized by a single out-of-plane three-fold rotational symmetry (C$_3$). It adopts the A$_2$BB'O$_6$ structural motif, where the A$_2$B sites correspond to three crystallographically distinct Ni positions, and the B' site corresponds to Te locations[18–20]. In Ni$_3$TeO$_6$, the Ni$^{2+}$ and Te$^{6+}$ ions reside within highly distorted oxygen octahedra, and based on the local octahedral environments, the Ni sites are categorized as Ni$^I$, Ni$^{II}$, and Ni$^{III}$. Vertical displacements of the Ni$^{2+}$ and Te$^{6+}$ ions away from the centers of their respective oxygen cages generate a net polarization along the *c*-axis, which can be either upwards (+) or downwards (–) as illustrated in Fig. 1a. The distortions of the oxygen octahedra further give rise to unequal O-O bond lengths in the triangular interfaces between the Ni$^I$/Ni$^{II}$O$_6$ and Ni$^{III}$/TeO$_6$ layers, which can be categorized as longest (red), intermediate (black), and shortest (blue) as marked in Fig. 1a. The orientation of these triangles, viewed along the *c*-axis, rotates either clockwise or counterclockwise, leading to a chiral structure with either left-handed (L) or right-handed (R) character. In an as-grown Ni$_3$TeO$_6$ crystal, polarity and chirality are reported to be interlocked[20], yielding only L+ and R– domain states (or alternatively, only L– and R+).

To see the origin of the interlocked relationship between polarity and chirality, we trace the spontaneous symmetry-breaking pathways from high-symmetry parent point groups to the polar-chiral point group 3. The parent groups that forbid both polar and chiral orders are either $\bar{3}$m or $\bar{6}$m2. Each of these two point groups can evolve into point group 3 via one of three distinct symmetry-breaking pathways, each involving a two-step process, as illustrated in Fig. 1b. In the high-symmetry parent phase ($\bar{3}$m or $\bar{6}$m2), all three order parameters vanish: the polar order $\langle p_z \rangle$, chiral order $\langle C \rangle$, and ferro-rotational order $\langle \varphi \rangle$ are all zero. In the first symmetry-breaking step, one of these three orders develops: a polar order ($\langle p_z \rangle \neq 0$) leads to point group 3m, a chiral order ($\langle C \rangle \neq 0$) yields 32, and a ferro-rotational order ($\langle \varphi \rangle \neq 0$) results in $\bar{3}$ (or $\bar{6}$) depending on the parent group. In the second step, the remaining two orders emerge in an interlocked fashion, yielding the final low-symmetry point group 3, characterized by coexisting, intertwined polar, chiral, ferro-rotational orders, $\langle p_z \rangle \neq 0$, $\langle C \rangle \neq 0$, and $\langle \varphi \rangle \neq 0$. For $Ni_3TeO_6$, where polarity and chirality are locked, we identify two possible symmetry-lowering pathways: $\bar{3}m \rightarrow \bar{3} \rightarrow 3$ and $\bar{6}m2 \rightarrow \bar{6} \rightarrow 3$, both of which involve the ferro-rotational order arising first (highlighted in red in Fig. 1b). Crucially, the distinction between these two routes lies in the symmetry operation that relates polar-chiral domain states: spatial inversion for the first path, versus a horizontal mirror plane for the second, as both can simultaneously reverse polarity and chirality in $Ni_3TeO_6$. At present, open questions remain as to which of these symmetry operations governs domain switching in $Ni_3TeO_6$, and what structural features characterize the corresponding domain walls.

To approach these open questions, we employ a multimodal optical platform, shown in Fig. 1c, to investigate as-grown, single-crystal $Ni_3TeO_6$, with emphasis on selecting a suite of targeted optical techniques to probe specific order parameters and their characteristics. Rotation anisotropy (RA) electric dipole/electric quadrupole second harmonic generation (ED/EQ SHG) is used to resolve the full symmetry of the polar/ferro-rotational order[12,14,16,21–25], while polarization-resolved transmission circular birefringence (tCB) is applied to detect the presence of chiral order[26,27]. In addition, wide-field ED/EQ SHG imaging distinguishes domains with opposite polarities/ferro-rotationalities[16,21,28,29], and scanning ED/EQ SHG and tCB microscopies are used to spatially resolve the polar/ferro-rotational and chiral orders, respectively, both within domains and across domain walls. Importantly, all optical modalities are integrated within a single experimental setup, enabling measurements to be performed on the same sample spot (up to a diffraction-limited resolution accuracy of 2 $\mu$m) under identical conditions (e.g., during the same thermal cycle, or under the same magnetic field sweep). This integrated, in-situ, multi-messenger optical approach is expected to provide a comprehensive characterization of materials with intertwined symmetry-breaking orders, for example, the intertwined polar, chiral, and ferro-rotational orders in $Ni_3TeO_6$.

**Spatial inversion relationship between polar-chiral domains in single crystalline $Ni_3TeO_6$**

We began our investigation with RA SHG measurements on a single crystal $Ni_3TeO_6$ to probe its full symmetry properties. Polar plots of RA SHG taken in the linearly parallel polarization channel, shown in Fig. 2a, correspond to three representative locations: two polar-chiral domains (labeled D1 and D2) and a domain wall (DW) between them. The RA SHG patterns from D1 and D2 appear nearly identical, exhibiting a clear $C_3$ rotational symmetry with a slight angular offset of approximately 7° from the crystallographic *a*- and *b*-axes. This observation is consistent with the polar-chiral point group 3 assigned to $Ni_3TeO_6$, as the lack of vertical mirrors and in-plane rotational axes in point group 3 naturally allows a rotation of the RA SHG lobes away from the in-plane crystal axes. Due to either the spatial inversion or horizontal mirror relationship between D1 and D2, their ED SHG susceptibility tensors are related by a $\pi$- or 0-phase shift, respectively, $\chi^{ED}_{ijk}(D1) = -\chi^{ED}_{ijk}(D2)$ or $\chi^{ED}_{ijk}(D1) =$ $+\chi_{ijk}^{\mathrm{ED}}(D2)$, with *i, j, k* = *x* and *y* for the normal incidence geometry, and thereby, their resulting SHG electric fields differ by a sign or remain the same at each polarization angle of the RA SHG measurement. However, because RA SHG detects the optical intensity, which is quadratic in the electric field and insensitive to the overall optical phase shift, the RA SHG patterns of D1 and D2 remain indistinguishable. In contrast, the RA SHG pattern measured at a domain wall, which is approximately parallel to the *a*-axis, shows a marked departure from the $C_3$ rotational symmetry. It features pronounced lobes near the wall direction, potentially suggesting the emergence of an in-plane polarization at domain walls to account for the broken $C_3$ symmetry.

To map the polar-chiral domain distribution across this $Ni_3TeO_6$ single crystal, we performed the wide-field SHG imaging measurement, which is sensitive to the optical phase difference between adjacent domains[21]. As shown in Fig. 2b, the wide-field SHG image reveals prominent dark lines of suppressed SHG intensity separating neighboring domains. These features arise from destructive interference between the SHG electric fields of domains D1 and D2, indicating a $\pi$-phase difference between their electric fields. This observation provides direct evidence that D1 and D2 are related by the spatial inversion operation, rather than by the horizontal mirror operation, pinning down the symmetry-breaking pathway of $\bar{3}$m → $\bar{3}$ → 3. Therefore, this wide-field SHG imaging not only shows the spatial distribution of structural domains in $Ni_3TeO_6$ single crystals, but also unambiguously establishes the spatial-inversion relationship between the two domain states with interlocked polarity and chirality.

**Enhanced polarity and suppressed chirality at domain walls in single crystalline $Ni_3TeO_6$**

To investigate the domain walls between polar chiral domains in $Ni_3TeO_6$, we performed scanning optical SHG microscopy, a technique specifically designed to minimize the destructive interference effect observed in wide-field SHG imaging. As shown in Fig. 2c, the scanning SHG image reveals a bright line with enhanced SHG intensities at the domain wall separating domains D1 and D2, in stark contrast to the corresponding dark line seen in Fig. 2b. This enhanced SHG signal at the domain wall typically suggests a local increase in the inversion-symmetry breaking. Additionally, linear optical microscopy images taken at both the fundamental and SHG wavelengths show a uniform appearance across the single crystal, with no observable contrast between domains and domain walls (Supplementary Information Note 1). This rules out variations in linear optical contrasts as the origin of the enhanced SHG signal. We therefore confidently attribute it to the presence of a stronger inversion-symmetry-breaking source, such as increased polarity or chirality localized at domain wall.

To pinpoint the source of this enhanced inversion symmetry breaking at the domain walls, we performed tCB measurements, which probes linear polarization rotation which is sensitive to chirality but insensitive to polarity. As shown in Fig. 2d, we conducted polarimetric measurements of the transmitted light through domains D1 and D2 and domain wall, using the same incident linear polarization oriented 45° clockwise from the *a*-axis. The transmitted light remains linearly polarized at all three sites. However, its polarization direction varies distinctly: in D1, the polarization rotates by +3° (counterclockwise); in D2, by –3° (clockwise); and crucially, at the domain wall, by 0° (no rotation). These results confirm the opposite chirality between D1 and D2, and unexpectedly, reveal the achiral nature of domain walls, despite the enhanced SHG signal observed at domain walls.

We further investigated the chiral domains and domain walls using scanning tCB microscopy. To enhance the contrast between domains with opposite chirality, we set the analyzer for the transmitted light at 93° relative to the incident polarization, allowing the transmission for D1 but suppressing that for D2 and producing tCB image of this $Ni_3TeO_6$ crystal shown in Fig. 2e. The spatial distribution of D1 and D2 closely matches that observed in the wide-field SHG image in Fig. 2b, providing direct visualization of the interlocked relationship between polarity and chirality in $Ni_3TeO_6$. To confirm the achiral nature of the domain walls, we adjust the analyzer to be 90° from the incident polarization, a configuration that highlights regions with no polarization rotation. The resulting tCB image in Fig. 2f shows a dark line with zero polarization rotation aligned along the domain wall, consistent with the absence of chirality at domain walls. The contrast between suppressed tCB and enhanced SHG signals at domain walls strongly indicates the change of the relationship between polarity and chirality from that within domains.

**Symmetry evolution across polar domain walls in single crystalline $Ni_3TeO_6$**

To better quantify the characteristics of domain walls in $Ni_3TeO_6$, we analyzed the intensity profiles of the SHG and tCB signals along the direction normal to the domain wall orientation, as shown in Fig. 3a. These line profiles were obtained by averaging 6 linecuts extracted from the images in Figs. 2c and 2f, respectively. While approaching the center of the domain wall, the tCB intensity, proportional to the optical rotation, gradually decreases, while the SHG intensity in the parallel channel increases markedly. These two traces exhibit clear opposite trends. Given that tCB signal originates exclusively from chirality and approaches zero at the domain walls, the enhanced SHG signal observed in the same region is attributed primarily to polarity. The contrasting behavior between tCB and SHG suggests a scenario that the polarization is suppressed along the out-of-plane direction, which leads to the suppression in chirality and tCB, but is enhanced in the in-plane directions, which contributes to the enhancement of SHG. Additionally, the SHG line profile reveals a polar domain wall width of 4.6µm, while the tCB trace shows an achiral domain wall width of 40µm, nearly an order of magnitude larger. This pronounced disparity in the domain wall widths for the two order parameters indicates that the polar order is the leading order parameter, while the chiral order is secondary.

In addition to the SHG intensity enhancement near the domain walls, the RA SHG patterns exhibit distinct evolutions across the domain wall, as shown in Fig. 3b. In the upper panel, the RA patterns are normalized to the maximum intensity at the center of the domain wall (indicated by the black arrow in the middle polar plot), highlighting the intensity evolution across the domain wall. The lower panel displays the same set of RA SHG patterns, but with each normalized to its own maximum intensity to more clearly illustrate changes in symmetry. As one approaches the center of the domain wall, the RA SHG pattern evolves from six even lobes, indicative of the out-of-plane $C_3$ rotational symmetry within the domains, to three asymmetric lobe pairs of varying size, signifying the breaking of the $C_3$ symmetry at the domain wall. Since the out-of-plane polarization naturally preserves the $C_3$ symmetry, this C₃ symmetry breaking is therefore attributed to the emergence of in-plane polarization components at the domain wall, echoing the simultaneous suppression of out-of-plane polarization but emergence of in-plane polarization indicated by the contrasting trend of SHG and tCB signals across the domain wall.

**Mixed Bloch–Néel type domain walls in single crystalline Ni₃TeO₆**

Conventional polar domain walls are classified as either Bloch-type or Néel-type, depending on the orientation of the in-plane polarization relative to the domain wall: in-plane polarization is parallel to Bloch-type walls, while it is perpendicular in Néel-type walls. To determine the domain wall type in Ni₃TeO₆, we employ Ginzburg–Landau theory informed by experimental observations.

The intertwined nature of polar, chiral, and ferro-rotational orders in Ni₃TeO₆ mandates a symmetry-allowed trilinear coupling term in the free energy, $F = \cdots + g p_z C \varphi + \cdots$, where $g$ is the coupling constant. Given that Ni₃TeO₆ follows the symmetry-breaking pathway $\bar{3}m \to \bar{3} \to 3$, the ferro-rotational order ($\varphi$), which emerges first and sets the stage for the interlocked polarity and chirality, is the rigid order parameter – its modulation across the crystal incurs the largest energy cost among the three. As a result, $\varphi$ remains uniform even across domain walls, allowing the trilinear coupling to be reduced to an effective bilinear form, $F = \cdots + g_0 p_z C + \cdots$, where $g_0 = g\langle\varphi\rangle$ with $\langle\varphi\rangle$ representing the spatially uniform ferro-rotational background. This effective coupling locks the relative signs of $p_z$ and $C$, leading to (R+, L–) domains for $g_0 < 0$ and (R–, L+) for $g_0 > 0$.

Experimental SHG measurements reveal strong signals across domain walls, implying that polar order persists at the wall. We therefore model the system with an instability toward a polar order and calculate it with the sine-Gordon theory, under the ferro-rotational point group $\bar{3}$ for a polarization vector $\vec{p} = (p_x, p_y, p_z)$. The resulting analytical soliton solutions (Supplementary Information Note 2) describe the spatial evolution of $p_x$, $p_y$, and $p_z$ across the domain wall. As shown in Fig. 4a, $p_z$ switches sign across the wall and vanishes at its center, while $p_x$, $p_y$, and their combined in-plane magnitude $p_\parallel = \sqrt{p_x^2 + p_y^2}$ all peak at the center. Here, $p_x$, perpendicular to the wall, corresponds to the Néel-type, while $p_y$, parallel to the wall, corresponds to the Bloch-type. Their coexistence reveals that the domain wall is of mixed Bloch–Néel character. Moreover, the ratio, $p_x/p_y$, remains constant across the domain wall, indicating that the in-plane polarization vector maintains a uniform orientation, though not aligned with crystallographic axes (Fig. 4a inset). This is consistent with RA-SHG patterns, which exhibit intensity variation but preserve orientation when crossing the domain wall.

Using the spatial profile of $p_z$, we then calculate the chiral order parameter profile $C$ by minimizing the free energy $F_C + F_{p_z-C}$, where $F_C$ is the intrinsic chiral order energy and $F_{p_z-C} = g_0 p_z C$ for the bilinear coupling (Supplementary Information Note 2). As shown in Figure 4b, the location dependence of $C$ closely follows that of $p_z$, but exhibits a broader domain wall width. This confirms the interlocked nature of $p_z$ and $C$ even at domain walls, where both approach zero in contrast to their finite values inside domains. The broader width of the chiral domain wall further suggests that $p_z$ acts as the leading order parameter while $C$ forms the secondary order parameter.

**Conclusion**

In summary, we have demonstrated that the intertwining of polar, chiral, and ferro-rotational orders in $Ni_3TeO_6$ plays a defining role in shaping both domain structures and domain wall characteristics. Crucially, the presence of a uniform ferro-rotational background across domains and domain walls underpins the interlocked relationship between polarity and chirality within domains and gives rise to mixed Néel- and Bloch-type domain walls. Given the nascent stage of research into disentangling these intertwined orders, complementary experimental techniques are essential for developing a more in-depth and comprehensive understanding. Nonetheless, our findings provide key insights into strategies for domain and domain wall control in systems exhibiting such intertwined orders. For instance, to alter the interlocking between polar and chiral orders, it is necessary to first switch the underlying ferro-rotational state; likewise, tuning domain wall properties or driving domain wall motion requires access to and manipulation of this foundational ferro-rotational order. Furthermore, this framework of intertwined orders can be extended to the magnetic analog, where ferromagnetic, helimagnetic, and ferro-toroidal[28,30] orders similarly form a closed and interdependent set.

**Methods**

**$Ni_3TeO_6$ single crystal growth**

$Ni_3TeO_6$ single crystals were grown by chemical vapor transfer method. The powder samples of $Ni_3TeO_6$ were synthesized first using the solid state reaction method. High-purity powder of NiO (99.99%) and $Te_2O_5$ (99.99%) were mixed in stoichiometric ratios. The mixed powders were ground, pelletized, and then sintered in air at 700°C for 72 hours in a box furnace, with two intermediate grinding. The resulting green powder samples were sealed in evacuated quartz tubes together with $Cl_2$ gas as a transport agent. The tubes were placed in a two-zone furnace for 7 days, where the hot and cold ends were maintained at 700°C and 600°C, respectively.

**Multimodal optical characterization platform**

Optical measurements, including SHG and tCB, were conducted within the multimodal optical platform shown in the schematic in Fig. 1c.

SHG measurements were conducted using a 200kHz pulsed laser with the center wavelength at 800nm and a pulse duration of 50 fs. For RA SHG and scanning SHG microscopy measurements, the laser was focused on the sample surface to a spot size of 5μm full width at half maximum (FWHM), while a galvo mirror was used in conjunction with a 4f-imaging system to adjust the position of the laser within the focal plane while maintaining isotropic fluence. Wavelength-appropriate half waveplates and polarizers were used to define the incident and reflected polarizations. Reflected

SHG light (400nm) was collected by a photomultiplier tube after a set of lowpass and bandpass filters to reject the fundamental wavelength. For wide-field imaging, the incoming fundamental beam divergence is adjusted so that a large area of the sample is illuminated. The SHG light is mapped onto a two-dimensional electron-multiplying charge-coupled device (EMCCD) through the imaging system.

tCB measurements were conducted using a supercontinuum white light fiber laser equipped with an acousto-optic tunable filter set to a center wavelength of 517nm. The incident laser was focused on the sample to a spot size of 7μm FWHM, and the transmitted light was then focused onto an amplified Si photodetector after a set of bandpass filters to isolate the 517nm wavelength.

**X-ray diffraction measurements**

X-ray diffraction was used to determine the crystal axes of the macroscopic specimen. Using a Rigaku SmartLab x-ray diffractometer, with a Cu-Kα source (wavelength: 1.5406 Å) and a Ge(220) monochromator, operated in the parallel-beam configuration, 2θ-ω scans (2θ: diffracted angle between the incident beam and the detector, ω: incident angle) were used to evaluate the lattice constant, diffraction angle, and orientation of the top facet. A phi scan (in-plane rotation) with in-plane stage rotation was used to determine the remaining crystal directions.

**Data availability**

All data supporting this work are available from the corresponding author upon request.

**Acknowledgements**

L.Z., K.S., and J.T.H. acknowledges the support from the National Science Foundation (NSF) through the Materials Research Science and Engineering Center at the University of Michigan, Award No. DMR-2309029. The work at Rutgers was supported by the W. M. Keck foundation grant to the Keck Center for Quantum Magnetism at Rutgers University. The work at the New Jersey Institute of Technology was supported by the NSF CAREER grant no. DMR-2236543.

**Author contributions**

W.Z., Y.A., and L.Z. conceived the idea and initiated the project. W.Z., Y.A., and X.G. performed the optical experiment, under the supervision of L.Z. J.Y. and S.-W. C. grew the $Ni_3TeO_6$ single crystals. T.C. performed the x-ray diffraction measurements to determine the crystal axes, under the supervision of J.T.H. N.A. performed the transmission electron microscopy measurements. K.S. performed the Ginzburg-Landau theory calculations. W.Z., J.H.Y, Y.A., and L.Z. analyzed the data and wrote the manuscript. All authors participated in discussion of the results.

**Competing interests**

The authors declare no competing interests.

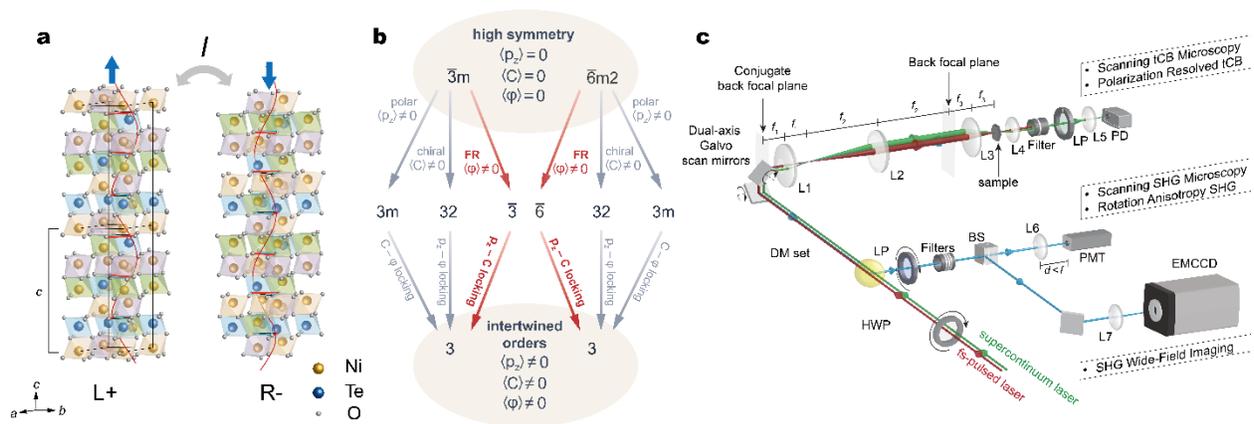

**Figure 1 | Crystal structure, symmetry-breaking pathways, and multimodal optical platform. a**, Crystal structure of $Ni_3TeO_6$, showing two domain states related by the spatial inversion (*I*) operation, each with interlocked polar and chiral orders. Polarity is shown by the blue arrows, while chirality is indicated by the rotation of the triangular interfaces between oxygen cages following the red helices. **b**, Symmetry-breaking pathways from two high-symmetry point group candidates where the polar ($p_z$), chiral (C), and ferro-rotational (φ) orders are zero, to the low-symmetry point group where each of the three order parameters are nonzero. The intermediate point groups each only contain one of the three order parameters, while the second symmetry-breaking step has the remaining two order parameters emerge locked by the extant order parameter. **c**, Schematic of the multimodal optical platform. Scanning optical SHG microscopy, RA SHG, wide-field SHG imaging, scanning tCB microscopy, and polarization resolved tCB measurements can be performed in the same setup by adjusting corresponding optical components to suitable wavelengths based on the laser source used for the measurements.

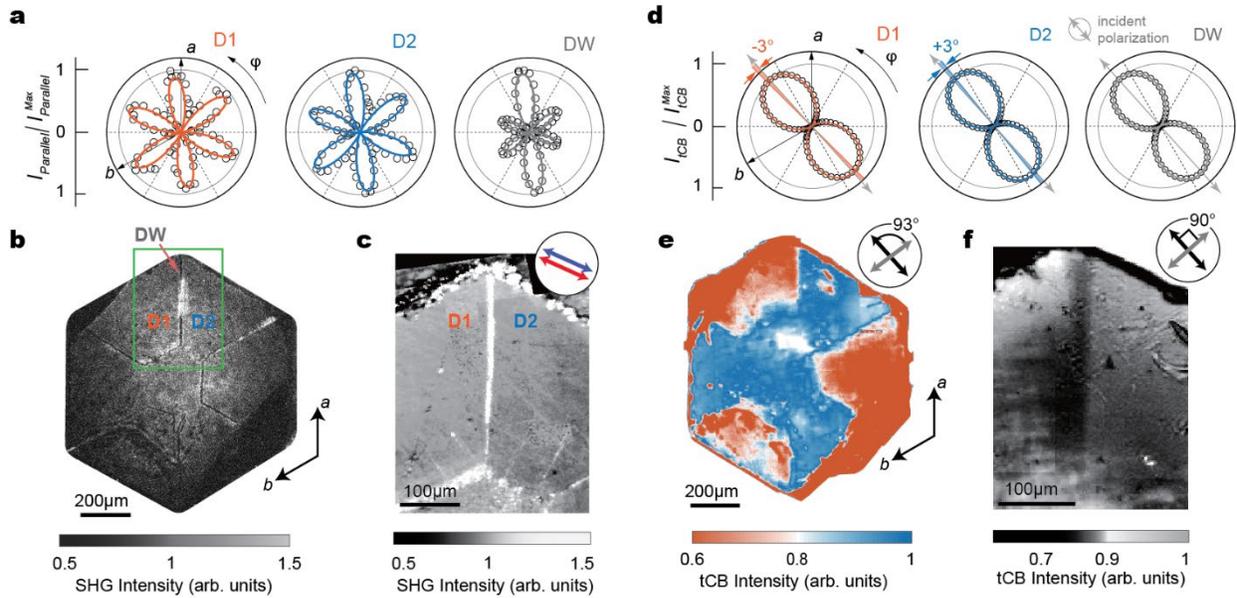

**Figure 2 | Full symmetry properties of Ni$_3$TeO$_6$ probed by the combination of SHG and tCB. a,** Polar plots of RA SHG taken in the parallel channel under the normal incidence at Domain 1 (D1), Domain 2 (D2), and the domain wall (DW), as shown in **b**. The analyzer for the reflected SHG (400nm) beam is co-rotated with the polarization of the incident fundamental (800nm) beam. Black open circles are experimental data, and colored curves are their best fit with the derived expression for ED SHG under the point group 3. **b,** Wide-field SHG image and **c,** Scanning SHG microscopy image captured with a fixed incident fundamental polarization and reflected SHG polarization as shown in the inset of **c**. In **c**, the polarization is chosen to maximize the measured SHG intensity. **d,** Polar plots of angle-resolved tCB at D1, D2, and DW, with the incident polarization marked as a double-sided arrow. The colored curves are the best fit with polarization rotation, which gives the polarization rotations as marked in the figure by the red and blue shaded areas. Scanning tCB microscopy images captured with the analyzer angle set **e,** at 93° and **f,** 90° from the incident polarization. The angle for **e** is chosen to selectively suppress the transmission at D2, while **f** is chosen to allow similar transmission between both domains.

13 / 15

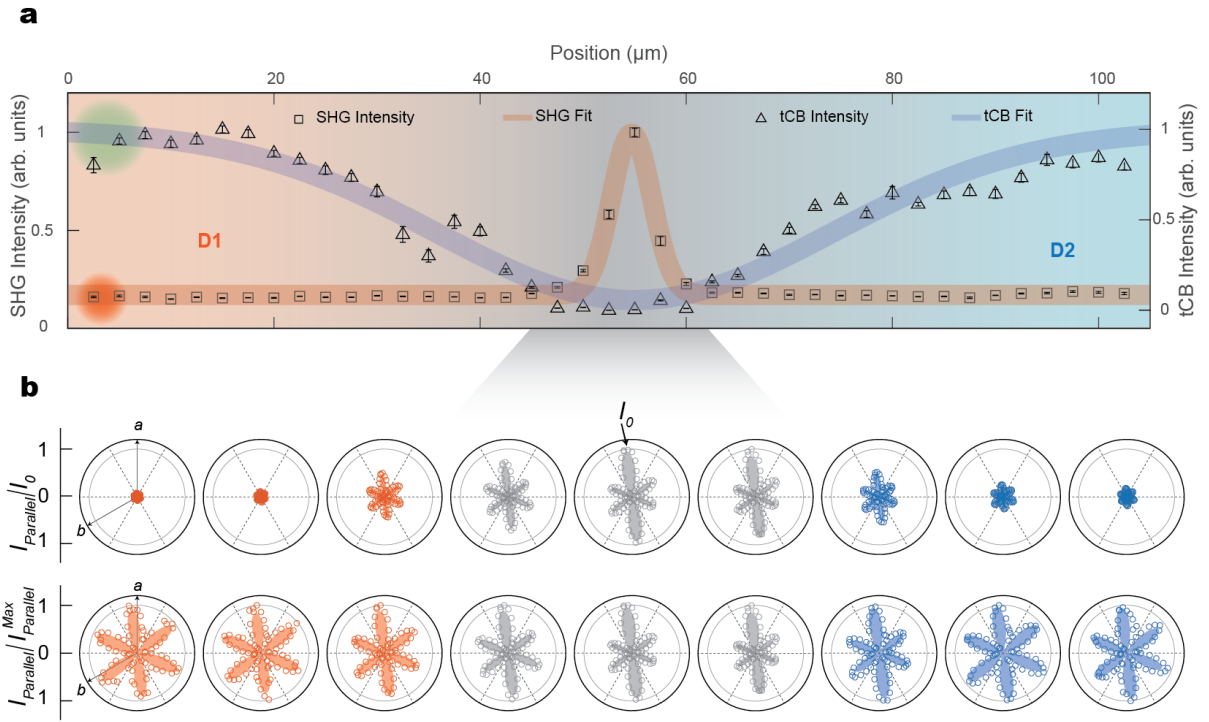

**Figure 3 | Evolution of polar and chiral orders across the domain wall in Ni$_3$TeO$_6$. a,** Spatial profile of SHG and tCB intensity across the domain wall as shown in Fig. 2c and 2f. Open squares and triangles are experimental data, while the colored curves are the Gaussian fits for the normalized intensity. The red and green spots indicate the spot size of the laser beam with respect to the scale of the *x* axis. **b,** Evolution of RA SHG across the domain wall. Open circles are experimental data, and shaded areas are their best fit with the derived expression for ED SHG based on symmetry. Top row is normalized by the maximal SHG intensity at the domain wall, and the polar plots on the bottom row are normalized with respect to the maximal SHG intensity at each position.

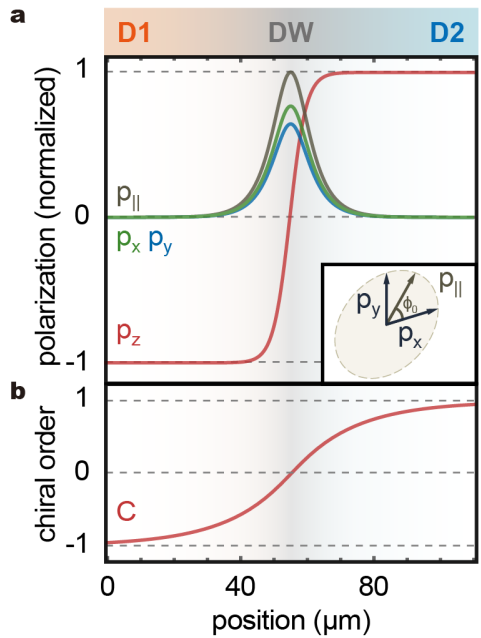

**Figure 4 | Mixed Bloch-Néel type domain wall shown by the G-L theory. a,** Computed sine-Gordon solution (normalized) for the minimization of Ginzburg-Landau free energy for the polar order parameter components $p_x$, $p_y$, and $p_z$ across the domain wall, as well as the in-plane component $p_{//} = \sqrt{(p_x^2 + p_y^2)}$. As shown in the inset, $p_x$ and $p_y$ are related by a control parameter $\phi_0$ which is not constrained by symmetry; subsequently, the in-plane polar component is generally nonzero in both the $x$ (Néel) and $y$ (Bloch) directions, leading to a mixed Bloch-Néel character. **b,** Chiral order (C) parameter profile over the domain wall, given by the sine-Gordon solution.

Supplementary Information for

# Intertwined polar, chiral, and ferro-rotational orders in a rotation-only insulator

Weizhe Zhang[1], June Ho Yeo[1], Xiaoyu Guo[1], Tony Chiang[2], Robert Hovden[2], John T. Heron[2], Kai Sun[1], Junjie Yang[3], Sang-Wook Cheong[4], Youngjun Ahn[1,5,+], Liuyan Zhao[1,+]

[1]Department of Physics, University of Michigan, Ann Arbor, MI, USA
[2]Department of Materials Science and Engineering, University of Michigan, Ann Arbor, MI, USA
[3]Department of Physics, New Jersey Institute of Technology, Newark, NJ, USA
[4]Keck Center for Quantum Magnetism and Department of Physics and Astronomy, Rutgers Center for Emergent Materials, Rutgers University, Piscataway, NJ, USA
[5]Materials Science Division, Argonne National Laboratory, Lemont, IL, USA

[+]Corresponding author: Liuyan Zhao (lyzhao@umich.edu) & Youngjun Ahn (youngjun.ahn@anl.gov)

**Table of Contents**

**Supplementary Note 1. Optical images of Ni₃TeO₆ crystal**

To clarify the origins of the domain walls visible in both the scanning SHG image and the SHG wide-field image, we performed a scanning reflectivity image across the region in which the domain walls are visible under SHG measurement. In contrast to Figure S1b, no contrast can be observed at the domain walls in the scanning reflectivity measurement, ruling out the role of linear optical contrast in the variation of the SHG signal at the domain wall.

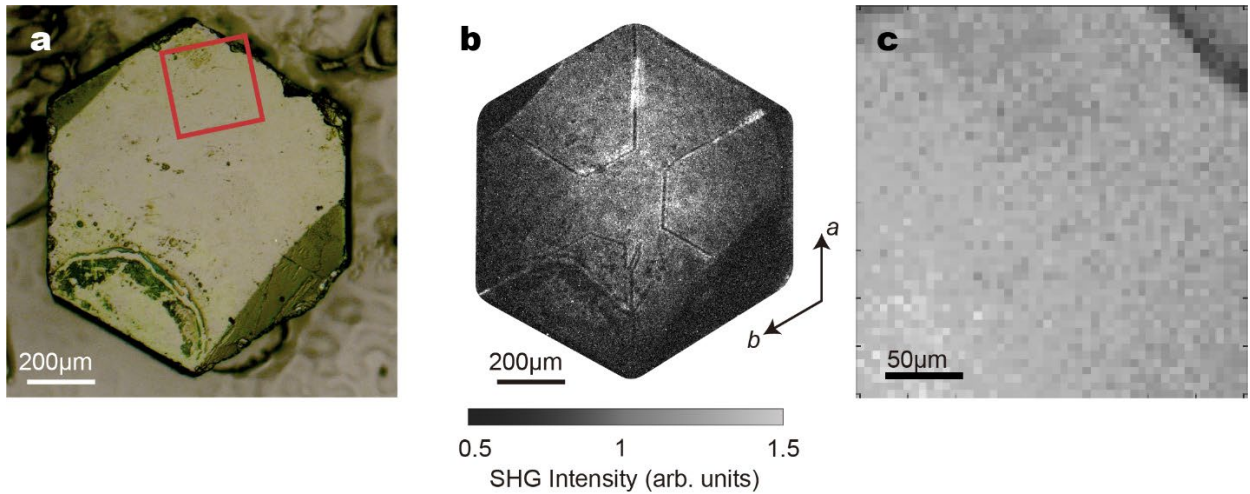

**Supplementary Figure 1. a,** Optical image of the Ni₃TeO₆ crystal as observed under an optical microscope. The area marked in red corresponds to the area on which the scanning image in **c** is taken. **b,** SHG wide-field image captured in normal incident under parallel HWP and LP orientations as shown in Fig. 2b of the main text, shown here for comparison. **c,** Scanning reflectivity captured with 800nm for the red area marked in **a**.

**Supplementary Note 2. Ginzburg-Landau theory and domain wall structures**

**2.1 Point-group hierarchy and symmetry-breaking pathways for the generation of polar and chiral order**

To understand the symmetry properties and interplay between polar and chiral orders, we begin with a high-symmetry crystal structure—referred to as the *parent state*—that forbids both types of order. Via spontaneous symmetry breaking, this high-symmetry state is reduced to a lower-symmetry configuration (point group 3), in which both polar and chiral orders can emerge.

In group-theoretical terms, the symmetry group of the parent state must contain point group 3 as a subgroup, while also including symmetry operations that forbid the existence of polar and chiral orders. By examining the hierarchy of crystallographic point groups, we find that there are exactly two such parent groups that satisfy these conditions.

In the first scenario (left), the parent state has point group symmetry $\bar{3}m$ ($D_{3d}$). In the second scenario (right), the parent state belongs to point group $\bar{6}m2$ ($D_{3h}$). In principle, one could also consider higher-symmetry parent states such as $6/mmm$ ($D_{6h}$), but in those cases, reducing to point group 3 necessarily requires additional symmetry breaking beyond polar, chiral, or ferro-rotational orders. For this reason, we restrict our discussion to the two cases summarized in Fig. 1b.

It is worthwhile emphasizing that for both parent states $\bar{3}m$ and $\bar{6}m2$, symmetry breaking to the lower symmetry group 3 requires two steps (two symmetry breakings) as shown in Fig. 1b. More importantly, this symmetry breaking down to point group 3 necessarily involves three distinct but intertwined order parameters: the polar order $p_z$, the chiral order $C$, and a ferro-rotational order $\varphi$. In contrast to $p_z$ and $C$, which are odd under spatial inversion, $\varphi$ preserves inversion symmetry but breaks vertical mirror symmetry.

These three order parameters $p_z$, $C$, and $\varphi$ characterize different symmetry-breaking patterns. In group-theory language, they belong to three distinct one-dimensional irreducible representations of the parent symmetry group. For the $\bar{3}m$ parent state, $p_z$ transforms as $A_{2u}$, and $C$ as $A_{1u}$; both are odd under spatial inversion. In contrast, $\varphi$ transforms as $A_{2g}$, which is even under inversion but odd under vertical mirror reflection. For the $\bar{6}m2$ case, $p_z$ and $C$ belong to the $A_2''$ and $A_1''$ representations, respectively—both odd under the horizontal mirror symmetry $\sigma_h$—while $\varphi$ transforms as $A_2'$, which is even under $\sigma_h$.

Remarkably, these three order parameters exhibit a nontrivial trifold *intertwining* relationship: the product of any two transforms identically to the third. For example, the product $p_z C$ transforms as $\varphi$; similarly, $p_z \varphi$ transforms as $C$, and $C\varphi$ transforms as $p_z$. This implies that if any two of the three order parameters acquire nonzero expectation values, the third is necessarily induced by symmetry.

This intertwined structure also explains why the symmetry-breaking transition from the parent state to point group 3 proceeds in two steps. In the high-symmetry parent state, all three order parameters vanish: $\langle p_z \rangle = \langle C \rangle = \langle \varphi \rangle = 0$. In the low-symmetry phase with point group 3, all three acquire nonzero expectation values. However, only two independent symmetry-breaking steps are required—each generating one of the three order parameters. The third is then automatically induced. Depending on the order of symmetry breaking, there are three distinct pathways from each parent state to the final low-symmetry phase:

1. **Polar-first pathway**: First generate polar order $p_z$, reducing the symmetry to an intermediate group $3m$ ($C_{3v}$). A second symmetry breaking then induces either $C$ or $\varphi$, with the third arising automatically.
2. **Chiral-first pathway**: First generate chiral order $C$, reducing the symmetry to another intermediate group 32 ($D_3$). A second step then produces $p_z$ or $\varphi$.
3. **Ferro-rotational-first pathway**: Begin by generating ferro-rotational order $\varphi$, reducing the symmetry to $\bar{3}$ ($C_{3i}$) or $\bar{6}$ ($C_{3h}$) with subsequent symmetry breaking producing $p_z$ or $C$

**2.2 Symmetry-breaking pathways and their connection to domain wall structures**

Although all three pathways lead to the same final symmetry (point group 3), they represent distinct physical behavior when order-parameter domains and domain walls are considered. In the low-symmetry phase where all three order parameters are present, a natural question arises: which order parameter is the most *rigid*—that is, which one incurs the highest energetic cost when forming a spatial modulation or domain wall? We refer to this as the **rigid order parameter**. Because domain wall formation in this parameter is energetically unfavorable, the system tends to keep it uniform across space, allowing the other two order parameters to vary more freely.

There are three possible scenarios: $p_z$, $C$, or $\varphi$ serving as the rigid order parameter. Each scenario corresponds to one of the three symmetry-breaking pathways mentioned above. This connection arises because a rigid order parameter, held uniform in space, effectively constrains the symmetry of the domain wall configuration to that of the corresponding intermediate state in the hierarchy.

To make this precise, we write down the Ginzburg–Landau free energy involving the three intertwined order parameters:

$$F = \iint dx\, dy \left[ \frac{|\nabla p_z|^2}{2m_z} + \frac{r_z}{2} p_z^2 + \frac{|\nabla C|^2}{2m_C} + \frac{r_C}{2} C^2 + \frac{|\nabla \varphi|^2}{2m_\varphi} + \frac{r_\varphi}{2} \varphi^2 + g\, p_z\, C\, \varphi \right.$$
$$\left. + \text{ higher order terms} \right],$$

Here, we showed the leading order terms up to cubic order, and the most important term here is the cubic coupling: $g\, p_z\, C\, \varphi$, which is allowed by symmetry. If one order parameter is rigid and spatially uniform—say, $\varphi$ —the cubic term reduces to an effective bilinear coupling:

$$f = \cdots + g_0\, p_z\, C\, + \cdots,$$

with $g_0 = g \langle \varphi \rangle$. This term reduces the symmetry of the Ginzburg–Landau free energy from that of the high-symmetry parent state to an intermediate symmetry group— $\bar{3}$ ($C_{3i}$) or $\bar{6}$ ($C_{3h}$). In doing so, this coupling term locks the relative signs of the other two order parameters. If $g_0 < 0$, then $p_z$ and $C$ prefer to have the same sign to minimize the free energy. Conversely, if $g_0 > 0$, the signs are opposite. In both cases, if the sign of $p_z$ flips across a domain boundary, the sign of $C$ will also flip in order to minimize the free energy.

The experimental observation that $p_z$ and $C$ are locked in Ni$_3$TeO$_6$ strongly suggests that $\varphi$ is the most rigid order parameter in this system. This places Ni$_3$TeO$_6$ within a specific symmetry-breaking pathway, where the ferro-rotational order acts as a rigid background, controlling the domain wall behavior of the polar and chiral orders.

### 2.3 Parent states

As discussed above, there are two possible parent states: $\bar{3}m$ or $\bar{6}m2$. If $p_z$ or $C$ is the most rigid order parameter—corresponding to the "polar-first" or "chiral-first" symmetry-breaking pathways—then it is not necessary to distinguish between these two parent states, because they share the same intermediate symmetry groups.

However, when $\varphi$ is the most rigid order parameter, the distinction between the two parent states becomes essential, as they lead to different intermediate symmetry groups, $\bar{3}$ or $\bar{6}$, and thus different symmetry constraints for domain walls.

If the parent state is $\bar{3}m$, the presence of a background $\langle\varphi\rangle \neq 0$ reduces the symmetry to the intermediate group $\bar{3}$. In this case, the two opposite order parameter domains of $p_z$ or $C$, $D_1$ and $D_2$, are related by spatial inversion. Since the ED-SHG tensor changes sign under inversion, the SHG signals from the two domains will exhibit a $\pi$ phase shift.

In contrast, if the parent state is $\bar{6}m2$, the intermediate symmetry is $\bar{6}$, and the two opposite domains are related by a horizontal mirror, which preserves the sign of the ED-SHG tensor. As a result, the SHG signals from domains $D_1$ and $D_2$ will be in phase, with no $\pi$ shift.

The experimental signature observed in Ni$_3$TeO$_6$ shows a $\pi$-phase shift between domains, indicating that the correct symmetry-breaking pathway corresponds to the first scenario—where the parent state has $\bar{3}m$ symmetry and the intermediate state is $\bar{3}$.

**2.4 Ginzburg–Landau theory for a domain wall**

With the symmetry properties identified experimentally, we can now construct the Ginzburg–Landau theory by including all symmetry-allowed terms.

Based on the experimental observation of locked polar and chiral orders, we assume that $\varphi$ (the ferro-rotational order) is the most rigid—meaning that in the crystal, it is energetically costly to form a domain wall of $\varphi$. As such, we treat $\varphi$ as a uniform background field with fixed sign, which does not change across the domain wall. As discussed earlier, the presence of $\varphi$ reduces the symmetry of the system from the high-symmetry parent group to an intermediate group—either $\bar{3}$ or $\bar{6}$. This intermediate symmetry will define the basis for constructing the Ginzburg–Landau free energy.

For Ni$_3$TeO$_6$, experimental signatures suggest that the intermediate symmetry is $\bar{3}$, which will be the focus of the calculations below. For comparison, we will also later present the Ginzburg–Landau theory for the case with $\bar{6}$ symmetry.

In addition to the rigid ferro-rotational background $\varphi$, we consider the following order parameters:

- The out-of-plane polar order $p_z$,
- The chiral order $C$,
- The in-plane components of the polar order: $p_x$ and $p_y$.

While $p_x$ and $p_y$ are forbidden in the uniform state by the threefold rotational symmetry, they can be induced near domain walls where this symmetry is locally broken. This will be demonstrated explicitly below.

The Ginzburg–Landau free energy takes the form:

$$F = \iint dx\, dy\, f(x, y)$$

where $f(x, y)$ is the free energy density at position $(x, y)$. Here, we assume the system is homogeneous along the $z$-direction, so only the $x$ and $y$ coordinates need to be considered. For $f(x, y)$, it takes the form:

$$f = f_{p_z} + f_C + f_{p_\parallel} + f_{p_z-C} + f_{p_z-p_\parallel} + \text{higher order terms}$$

where each term is defined as follows:

**Polar order $p_z$:**

$$f_{p_z} = \frac{|\nabla p_z|^2}{2m_z} + \frac{r_z}{2} p_z^2 + \frac{u}{4} p_z^4$$

Here, we assume $r_z < 0$, so that the quadratic term has a negative coefficient and drives an instability that generates a spontaneous polar order $p_z \neq 0$.

**Chiral order C:**

$$f_C = \frac{|\nabla C|^2}{2m_C} + \frac{r_C}{2} C^2$$

We assume $r_C > 0$, meaning that $C$ by itself does not spontaneously break symmetry. Instead, it is induced by the instability of $p_z$ through their couplings (see below). While it is also possible to take $C$ as the primary instability and generate $p_z$ secondarily, as will be shown below, this leads to the opposite trend in domain wall widths: if $p_z$ drives the instability, then its domain wall is narrower than that of $C$—a trend consistent with experimental observations. In contrast, if $C$ is the primary instability, it would predict a broader domain wall for $p_z$.

**In-plane polar components $p_x$ and $p_y$:**

$$f_{p_\parallel} = \frac{|\nabla p_x|^2 + |\nabla p_y|^2}{2m_\parallel} + \frac{A}{m_\parallel}\left[(\partial_x p_x)^2 - (\partial_y p_x)^2 - (\partial_x p_y)^2 + (\partial_y p_y)^2 + 4\partial_x p_x \partial_y p_y\right]$$
$$+ \frac{B}{m_\parallel}\left[\partial_x p_x \partial_y p_x - \partial_x p_y \partial_y p_y - \partial_x p_x \partial_x p_y + \partial_y p_x \partial_y p_y\right] + \frac{r_\parallel}{2}(p_x^2 + p_y^2).$$

We set the quadratic coefficient $r_\parallel > 0$, so that $\langle p_x \rangle = \langle p_y \rangle = 0$ in a uniform state. Here, we have also included all symmetry-allowed leading-order derivative terms.

**Coupling terms between $p_z$ and $C$:**

$$f_{p_z-C} = g_0 p_z\, C$$

As discussed above, this term arises due to the presence of a rigid-order-parameter background order $\langle \varphi \rangle \neq 0$. It locks the relative signs of $p_z$ and $C$, such that across a domain wall where $p_z$ changes sign, $C$ must also flip in order to minimize the free energy.

**Coupling terms between $p_z$ and $p_x$, $p_y$:**

At leading order, symmetry forbids direct quadratic couplings between $p_z$ and $p_x$, $p_y$. However, symmetry does allow coupling between $p_x$, $p_y$ and the second-order spatial derivatives of $p_z$:

$$f_{p_z-p_\parallel} = \alpha\left[p_x(\partial_x^2 - \partial_y^2)p_z + 2p_y \partial_x \partial_y p_z\right] + \beta\left[2p_x \partial_x \partial_y p_z + p_y(\partial_x^2 - \partial_y^2)p_z\right].$$

These terms vanish in a uniform state but become active near a domain wall where the second order derivatives of $p_z$ become nonzero. As a result, the spatial variation of $p_z$ induces nonzero $p_x$ and $p_y$.

**2.5 Domain wall configurations**

Using the Ginzburg–Landau free energy constructed above, we can introduce a domain wall and compute the spatial profiles of all relevant order parameters by minimizing the total free energy.

For simplicity, we adopt one approximation: we assume that the induced order parameters $C$, $p_x$, and $p_y$ are weak, so their feedback on the primary order parameter $p_z$ can be neglected. This assumption is not essential for the qualitative features discussed below, but it allows for analytic solutions that capture the key physics more transparently.

We begin with the primary order parameter $p_z$, whose free energy $f_{p_z}$ is given above. In a uniform system ($\nabla p_z = 0$), minimizing the Ginzburg–Landau free energy yields two degenerate ground states:

$$p_z = \pm\sqrt{|r_z|/u}.$$

Now we consider a domain wall separating two domains located at $x < x_0$ and $x > x_0$. Without loss of generality, we choose the domain wall normal to lie along the $x$-direction, and impose the boundary conditions:

$$p_z(x \to -\infty) = -\sqrt{r_z/u} \quad \text{and} \quad p_z(x \to +\infty) = +\sqrt{r_z/u}.$$

Minimizing the Ginzburg–Landau free energy with these boundary conditions leads to the following differential equation:

$$-\frac{\partial_x^2 p_z}{m_z} - r_z\, p_z + u\, p^3 = 0,$$

with the boundary conditions as stated above. This equation admits an analytic kink solution:

$$p_z(x) = \sqrt{\frac{r_z}{u}} \tanh\left(\frac{x - x_0}{\sqrt{2}\xi_z}\right),$$

where $x_0$ is the location of the domain wall and $\xi_z = 1/\sqrt{r_z m_z}$ is the correlation length of the polar order.

Having obtained the spatial profile of $p_z(x)$, we now turn to the chiral order parameter $C(x)$, governed by the terms

$$F_C + F_{p_z-C} = \iint dx\, dy\, [\frac{|\nabla C|^2}{2m_C} + \frac{r_C}{2} C^2 + g_0 p_z\, C]$$

Here, the coupling term $g_0 p_z\, C$ acts like an external field for $C$, proportional to $p_z$. In a uniform system with constant $p_z$ and $C$, the minimum of this energy yields:

$$C = -\frac{g_0}{r_c} p_z$$

demonstrating the locking of $C$ and $p_z$ due to the background ferro-rotational order $\varphi$.

In the presence of a domain wall, where $p_z(x)$ varies in space, $C(x)$ is subject to a spatially varying effective field. Minimizing the free energy leads to the following differential equation:

$$-\frac{\partial_x^2 C}{m_C} + r_C\, C + g_0\, p_z(x) = 0,$$

with boundary conditions:

$$C(x \to -\infty) = \frac{g_0}{r_c}\sqrt{\frac{r_z}{u}} \quad \text{and} \quad C(x \to +\infty) = -\frac{g_0}{r_c}\sqrt{\frac{r_z}{u}}.$$

This equation is analytically solvable, yielding a smooth domain-wall profile for $C(x)$ that mirrors and broadens relative to $p_z(x)$

$$C(x) = -\frac{g_0}{r_c}\sqrt{\frac{r_z}{u}}\Bigg[1 + \frac{\exp\left(\frac{x-x_0}{\xi_C}\right)}{\mathrm{sinc}\left(\frac{\xi_z}{\sqrt{2}\xi_C}\pi\right)} - {}_2F_1\left(1, -\frac{\xi_z}{\sqrt{2}\xi_C}; 1 - \frac{\xi_z}{\sqrt{2}\xi_C}; -e^{\frac{\sqrt{2}}{\xi_z}(x-x_0)}\right)$$

$$- {}_2F_1\left(1, \frac{\xi_z}{\sqrt{2}\xi_C}; 1 + \frac{\xi_z}{\sqrt{2}\xi_C}; -e^{\frac{\sqrt{2}}{\xi_z}(x-x_0)}\right)\Bigg].$$

where ${}_2F_1(a,b;c;x)$ is the hypergeometric function, $\mathrm{sinc}\, x = \frac{\sin x}{x}$, and $\xi_z = 1/\sqrt{r_z m_z}$ and $\xi_C = 1/\sqrt{r_C m_C}$ are the correlation lengths of the polar and chiral order respectively.

Finally, we turn to the in-plane polar components $p_x(x)$ and $p_y(x)$ governed by

$$F_{p_\parallel} + F_{p_z - p_\parallel} = \iint dx\, dy\, [f_{p_\parallel} + f_{p_z - p_\parallel}]$$

where $f_{p_\parallel}$ and $f_{p_z - p_\parallel}$ are defined above. Because the domain wall normal is set to lie along the $x$-direction, $\partial_y p_x = \partial_y p_y = \partial_y p_z = 0$, and therefore, we get

$$F_{p_\parallel} + F_{p_z-p_\parallel} = \iint dx\, dy \left[ \frac{1+A}{2m_\parallel}(\partial_x p_x)^2 + \frac{1-A}{2m_\parallel}(\partial_x p_y)^2 - \frac{B}{m_\parallel}\partial_x p_x \partial_x p_y + \frac{r_\parallel}{2}(p_x^2 + p_y^2) \right.$$
$$\left. + \alpha\, p_x \partial_x^2 p_z - \beta\, p_y \partial_x^2 p_z \right].$$

As seen from the last two terms, $\partial_x^2 p_z$ acts as an effective external field for $p_x$ and $p_y$. Near a domain wall, this coupling generates a localized field that induces nonzero $p_x(x)$ and $p_y(x)$. In the uniform domains far away from the wall, where $\partial_x^2 p_z = 0$, this effective field vanishes, and both $p_x$ and $p_y$ return to zero.

To find the profile of $p_x(x)$ and $p_y(x)$, we minimize the free energy $F_{p_\parallel} + F_{p_z-p_\parallel}$, which yields:

$$-\frac{1+A}{m_\parallel} \partial_x^2 p_x + \frac{B}{m_\parallel} \partial_x^2 p_y + r_\parallel\, p_x + \alpha\, \partial_x^2 p_z = 0$$

$$-\frac{1-A}{m_\parallel} \partial_x^2 p_y + \frac{B}{m_\parallel} \partial_x^2 p_x + r_\parallel\, p_y + \beta\, \partial_x^2 p_z = 0$$

and the boundary conditions

$$p_x(x \to \pm\infty) = 0 \quad \text{and} \quad p_y(x \to \pm\infty) = 0.$$

To simplify the solution, we set $B = 0$, which removes the transverse coupling between $p_x$ and $p_y$. This approximation does not affect the qualitative behavior, and enables a straightforward analytic solution:

$$p_x(x) = -\frac{\alpha}{r_\parallel} \sqrt{\frac{r_z}{u}} \frac{1}{(1+A)\xi_\parallel^2} \left[ 1 + \frac{\exp\left(\frac{x-x_0}{\sqrt{1+A}\,\xi_\parallel}\right)}{\text{sinc}\left(\frac{\pi\, \xi_z/\xi_\parallel}{\sqrt{2(1+A)}}\right)} - \tanh\left(\frac{x-x_0}{\sqrt{2}\xi_z}\right) \right.$$
$$\left. -\, _2F_1\left(1, -\frac{\xi_z/\xi_\parallel}{\sqrt{2(1+A)}}; 1 - \frac{\xi_z/\xi_\parallel}{\sqrt{2(1+A)}}; -e^{\frac{\sqrt{2}}{\xi_z}(x-x_0)}\right) \right.$$
$$\left. -\, _2F_1\left(1, +\frac{\xi_z/\xi_\parallel}{\sqrt{2(1+A)}}; 1 + \frac{\xi_z/\xi_\parallel}{\sqrt{2(1+A)}}; -e^{\frac{\sqrt{2}}{\xi_z}(x-x_0)}\right) \right]$$

and

$$p_y(x) = -\frac{\beta}{r_\parallel}\sqrt{\frac{r_z}{u}}\frac{1}{(1-A)\xi_\parallel^2}\left[1 + \frac{\exp\left(\frac{x-x_0}{\sqrt{1-A}\xi_\parallel}\right)}{\mathrm{sinc}\left(\frac{\pi\,\xi_z/\xi_\parallel}{\sqrt{2(1-A)}}\right)} - \tanh\left(\frac{x-x_0}{\sqrt{2}\xi_z}\right)\right.$$

$$\left. - {}_2F_1\left(1, -\frac{\xi_z/\xi_\parallel}{\sqrt{2(1-A)}}; 1 - \frac{\xi_z/\xi_\parallel}{\sqrt{2(1-A)}}; -e^{\frac{\sqrt{2}}{\xi_z}(x-x_0)}\right)\right.$$

$$\left. - {}_2F_1\left(1, +\frac{\xi_z/\xi_\parallel}{\sqrt{2(1-A)}}; 1 + \frac{\xi_z/\xi_\parallel}{\sqrt{2(1-A)}}; -e^{\frac{\sqrt{2}}{\xi_z}(x-x_0)}\right)\right]$$

where ${}_2F_1(a,b;c;x)$ denotes the hypergeometric function and $\mathrm{sinc}\,x = \frac{\sin x}{x}$. The correlation lengths are given by $\xi_z = 1/\sqrt{r_z m_z}$ for $p_z$ and $\xi_\parallel = 1/\sqrt{r_\parallel m_\parallel}$ for $p_x$ and $p_y$. In general, these solutions show that near a domain wall, both $\langle p_x \rangle \neq 0$ and $\langle p_y \rangle \neq 0$.

### 2.6 $\bar{6}m2$ parent states

If the parent state has $\bar{6}m2$ symmetry, the domain-wall theory will largely remain the same, with only one modification in the coupling between $p_z$ and $p_x, p_y$

$$f_{p_z-p_\parallel} = \alpha\left(p_x\partial_x p_z^2 + p_y\partial_y p_z^2\right) + \beta\left(p_x\partial_y p_z^2 - p_y\partial_x p_z^2\right) + \gamma_1\left[p_x(\partial_x^2 - \partial_y^2)p_z^2 + 2p_y\partial_x\partial_y p_z^2\right]$$
$$+ \delta_1\left[2p_x\partial_x\partial_y p_z^2 + p_y(\partial_x^2 - \partial_y^2)p_z^2\right]$$
$$+ \gamma_2\left[p_x(\partial_x p_z)^2 - p_x(\partial_y p_z)^2 + 2p_y\partial_x p_z\partial_y p_z\right]$$
$$+ \delta_2\left[2p_x\partial_x p_z\partial_y p_z + p_y(\partial_x p_z)^2 - p_y(\partial_y p_z)^2\right]$$

Here, it is the derivatives of $p_z^2$ —rather than $p_z$ itself— that act as an effective external field inducing $p_x$ and $p_y$. In general, such couplings also give rise to nonzero $p_x$ and $p_y$ near the domain wall. However, in contrast to the solution found for the $\bar{3}m$ case discussed above, where $p_x$ and $p_y$ are anti-symmetric across the domain wall, the domain-wall configuration here is more complex: $p_x$ and $p_y$ are generally neither symmetric nor antisymmetric, but instead exhibit a mixture of both. As a result, such a domain wall will generally exhibit a nonzero in-plane electric dipole moment.

## 2.7 Spontaneous Symmetry Breaking at the Domain Wall

As discussed in Section 2.5, when the two domains $D_1$ and $D_2$ are related by spatial inversion symmetry, the in-plane polarization components $p_x$ and $p_y$ are antisymmetric across the domain wall (i.e., they flip sign at the boundary).

In this section, we consider a more sophisticated scenario in which the domain wall **spontaneously breaks inversion symmetry**. We again assume the presence of a background ferro-rotational order, which reduces the symmetry of the system to $\bar{3}$. In addition, we assume that the system exhibits an instability toward ferroelectric ordering $\langle \boldsymbol{p} \rangle \neq \boldsymbol{0}$, with an easy-axis anisotropy favoring alignment along the $z$-direction.

Unlike the previous model where only $p_z$ was treated as the primary order parameter, here we consider the situation in which ferroelectric ordering along *any* direction lowers the free energy. However, due to the easy-axis anisotropy, alignment along $+z$ or $-z$ is energetically preferred.

To model this easy-axis ferroelectric order, we employ the **sine-Gordon theory**. First, we express the 3D polarization vector in spherical (polar) coordinates:

$$(p_x, p_y, p_z) = p_0(\sin\theta\cos\phi, \sin\theta\sin\phi, \cos\theta).$$

The Ginzburg-Landau free energy can then be written as:

$$F = F_{p_0} + F_\theta + F_\phi$$

Here, $F_{p_0}$, $F_\theta$, and $F_\phi$ describe the amplitude, polar angle, and azimuthal angle degrees of freedom, respectively. Although coupling terms between these components exist, they are higher-order and can be neglected at leading order.

Because the amplitude mode $p_0$ (i.e., the Higgs mode) is typically gapped, we can, to leading order, neglect its fluctuations and treat it as a constant equal to its expectation value. The azimuthal angle $\phi$ determines the orientation of the in-plane component. To leading order, $\phi$ is governed by the free energy functional

$$F_\phi = \iint dx\, dy \left[ \frac{|\nabla\phi|^2}{2m_\phi} - \frac{r_\phi}{2} \cos 6\,(\phi - \phi_0) \right]$$

where the first term penalizes spatial inhomogeneity, favoring a homogeneous ferroelectric order. The second term, allowed by the $\bar{3}$ ($C_{3i}$) symmetry, energetically selects six preferred orientations $\phi = \phi_0 + n\pi/3$ with $n$ being an integer and $\phi_0$ is a control parameter. Without loss of generality, we assume here $r_\phi > 0$. Below, without loss of generality, we set the expectation value of $\phi$ to $\phi_0$, which minimizes the free energy, and assume it is spatially homogeneous, i.e., $\phi = \phi_0$ throughout the system.

It is also worth noting that the conclusion $\phi = \phi_0$ represents an approximation, based on the assumption that couplings between $\phi$ and other modes are negligible. In reality, such couplings can lead to spatial variations in $\phi$, an effect we will ignore in the present analysis for simplicity. As an example of such couplings, the $\bar{3}$ ($C_{3i}$) symmetry group allows an interaction term between the polar angle $\theta$ and the azimuthal angle $\phi$:

$$F_{\theta-\phi} = \iint dx\, dy\, \left[g_{\theta-\phi} \cos\theta \sin^3\theta \cos 3(\phi - \phi_1)\right].$$

Depending on the sign of $p_z = p_0 \cos\theta$, (i.e., whether $p_z > 0$ or $p_z < 0$), this term energetically favors in-plane orientations $\phi = \phi_1 + 2n\pi/3$ or $\phi = \phi_0 + \pi + 2n\pi/3$, respectively. Because, in general, $\phi_0 \neq \phi_1$, this coupling tends to tilt the in-plane polarization direction away from the orientation favored by the $\cos 6(\phi - \phi_0)$ term. Near the domain wall, the symmetry is further reduced, and additional couplings between $\phi$ and other modes can emerge, introducing further rotations of the energetically preferred in-plane direction. Nonetheless, even in the absence of such corrections, the low symmetry within each domain and at the domain boundary implies that the in-plane polarization direction is not expected to align with any crystallographic axis. Therefore, these additional perturbations do not qualitatively alter the conclusions and will be neglected in the following analysis.

For comparison, in systems with $\bar{6}$ ($C_{3h}$) symmetry, an alternative symmetry-allowed term of the form $\propto \cos 3(\phi - \phi_2)$ can appear. This term, unlike $F_{\theta-\phi}$, is independent of $\theta$ or $p_z$, and favors three orientations $\phi = \phi_2 + 2n\pi/3$. In contrast, the coupling term $F_{\theta-\phi}$ becomes symmetry-forbidden under the $\bar{6}$ symmetry.

The most important degree of freedom here is the **polar angle $\theta$**, which determines the out-of-plane direction of the polarization. The GL free energy for $\theta$ takes the form of a **sine-Gordon model**:

$$F_\theta = \iint dx\, dy\, \left[\frac{|\nabla\theta|^2}{2m_\theta} - \frac{r_\theta}{2}\cos 2\theta\right]$$

The first term penalizes spatial inhomogeneity, favoring uniform $\theta$, while the second term represents the easy-axis anisotropy. For $r_\theta > 0$, the potential is minimized at $\theta = 0$ or $\pi$, corresponding to polarization aligned along $+z$ or $-z$.

In the bulk (far from the domain wall), the ground state is either $\theta = 0$ or $\pi$, yielding two degenerate states with $p_z > 0$ and $p_z < 0$, and $p_x = p_y = 0$. Near the domain wall, minimizing the free energy leads to a domain-wall profile governed by a sine-Gordon differential equation.

$$-\frac{\nabla^2 \theta}{m_\theta} + r_\theta \sin 2\theta = 0$$

with boundary conditions

$$\theta(x \to -\infty) = \pi \quad \text{and} \quad \theta(x \to +\infty) = 0.$$

This equation admits an **analytical kink (soliton) solution**, which smoothly interpolates between $\theta = 0$ and $\theta = \pi$.

This implies that the $z$-component and in-plane components of the polarization vector behave as:

$$p_x(x) = p_0 \sin\theta(x) \cos\phi(x) = p_0 \,\text{sech}\frac{x - x_0}{\xi_\theta/\sqrt{2}} \cos\phi_0$$

$$p_y(x) = p_0 \sin\theta(x) \sin\phi(x) = p_0 \,\text{sech}\frac{x - x_0}{\xi_\theta/\sqrt{2}} \sin\phi_0$$

$$p_z(x) = p_0 \cos\theta(x) = p_0 \tanh\frac{x - x_0}{\xi_\theta/\sqrt{2}}$$

respectively. Here, the correlation length is $\xi_\theta = 1/\sqrt{r_\theta m_\theta}$. Notably, this sine-Gordon solution yields a $p_z(x)$ profile very similar to the earlier $p_z^4$-theory, where $p_z(x) = \sqrt{\frac{r_z}{u}}\tanh\left(\frac{x-x_0}{\sqrt{2}\xi_z}\right)$. However, the in-plane component $p_\parallel(x) = \sqrt{p_x^2 + p_y^2}$ now **peaks at the domain wall**, rather than vanishing there as it did under inversion symmetry. This peak arises because the ferroelectric order spontaneously breaks inversion symmetry at the domain boundary. In addition, since there is no symmetry that pins the value of $\phi_0$, its value is generally not constrained to high-symmetry directions, such as integer or half-integer multiples of $\pi$. As a result, unless fine-tuned, both $p_x$ and $p_y$ will be nonzero, leading to a mixed Néel–Bloch type domain wall.

As discussed earlier, $p_z$ acts as an effective external field for the chiral order $C$. Therefore, the spatial profile of $p_z$ imposes an inhomogeneous driving force on $C$, inducing a corresponding spatial modulation. In principle, the in-plane polarization $p_\parallel$ could also influence $C$, but its coupling is higher order (e.g., involving additional powers of $p_\parallel$ or spatial derivatives). Thus, at leading order, we can just focus on the coupling between $p_z$ and $C$.

Following a similar analysis as above, we can determine the profile of $C$ by solving the relevant differential equation, subject to the domain-wall boundary conditions. The resulting solution for $C$ is qualitatively similar to that found in the inversion-symmetric case, differing primarily by a rescaling of the domain-wall width

$$C(x) = -\frac{g_0 p_0}{r_c}\left[1 + \frac{\exp\left(\frac{x-x_0}{\xi_C}\right)}{\mathrm{sinc}\left(\frac{\xi_\theta}{2\sqrt{2}\xi_C}\pi\right)} - {}_2F_1\left(1, -\frac{\xi_\theta}{2\sqrt{2}\xi_C}; 1 - \frac{\xi_\theta}{2\sqrt{2}\xi_C}; -e^{\frac{2\sqrt{2}}{\xi_\theta}(x-x_0)}\right) \right.$$
$$\left. - {}_2F_1\left(1, \frac{\xi_\theta}{2\sqrt{2}\xi_C}; 1 + \frac{\xi_\theta}{2\sqrt{2}\xi_C}; -e^{\frac{2\sqrt{2}}{\xi_\theta}(x-x_0)}\right)\right].$$

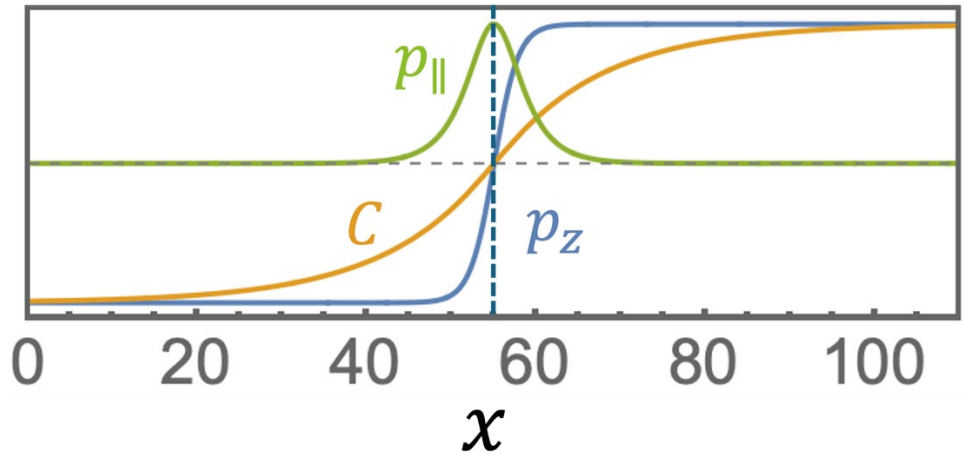

**Supplementary Figure 2.** Spatial profiles of order parameters across a domain wall. The plots show the spatial variation of the order parameters $p_z$, $C$, and $p_\parallel$ across a domain wall, calculated from the Sine-Gordon theory. The vertical dashed line marks the center of the domain wall. For this plot, we set the correlation lengths as follows: $\xi_\theta=4$ for the ferroelectric order and $\xi_C=12$ for the chiral order $C$. For both ferroelectric and chiral orders, the order parameter values are normalized by their respective expectation values in the bulk, far from the domain wall.